# Working Relations[1]

Steven J. Jackson
Professor of Information Science and Science and Technology Studies
Cornell University
110H Gates Hall, Ithaca, New York
USA 14850
e-mail: sjj54@cornell.edu
https://orcid.org/0000-0002-4426-1320

**Abstract**
This paper offers a concept of working relations as a complement and extension to existing theories of maintenance, care and repair. Building on the cases of an umbrella, a tractor and a pond, it advances seven propositions that might guide and inform further work and thinking in this space. It concludes with the challenging figures of Chernobyl, nickel extraction, and AI, and argues for the centrality of working relations to more generative and pluralistic relations with the things and worlds around us.

“I have forgotten my umbrella.”
(Unpublished note found among the papers of Friedrich Wilhelm Nietzsche upon his death on August 25, 1900)

## Introduction

The event[i] that occasioned these thoughts was a lovely and instructive week around Tartu and the Estonian National Museum in the company of many of the contributors to this volume. The stunning Estonian National Museum stands at the end of the decommissioned runway of one of the largest Soviet air bases of the Cold War, its curves built to mimic the takeoff path of the bombers that never, in the end, flew west. Our common thread was repair, and the distinctive forms of mending, reuse, and everyday survival that characterized life in Estonia in the late Soviet era. I learned the miracles of machines kept going long past Western shelf lives; of bicycles remade into harvesters and musical instruments; of the 124 (and counting!) known uses of newspaper (see also Errázuriz and Greene 2021); of local food systems and cousins in the country that kept cities fed as shelves emptied amidst a famine that never was. Participants and hosts alike brought a fantastic array of sites and stories to the table, from invisible histories and presents of craft and mending, to reimaginations of building care, to unwasted fashion and textiles, to new directions in degrowth. To this I wish to add three more.

[1] The author wishes to thanks Geof Bowker, Matt Ratto, Matt Stahl, Bill Gaver, members of the Computing On Earth Lab, and the anonymous reviewers of this volume for feedback on prior versions of this paper. As always, the bad ideas remain his own.

*The first*: a lovely day spent learning to fix umbrellas with an acclaimed Parisian master of the craft. The context was a workshop of artists and repair scholars organized by Dutch colleagues. We slowly and imperfectly learned the art of spindle cutting and replacement, the tensioning of the delicate and amazing apparatus that allows umbrellas to expand and retract, the art of threading and reattaching skeleton and skin. At the conclusion of our lesson we were given a stock of umbrellas broken in myriad ways and invited to do what we liked. Most present turned their umbrellas into a series of fantastic and imaginative objects; with their graceful mechanics, insect-like strength and delicacy, and rich symbolics – weather, water, childhood sheltering – umbrellas are surely evocative objects par excellence.

I however wanted my umbrella to be… an umbrella. I picked a not *too* broken option from the pile – a plain black single fold with a worn wooden handle. The fabric had come unhooked from the mechanics on one side, torn by the force of an unknown storm in an unknown city. While black thread was on hand, I selected instead a deep orange, and worked slowly to reattach fabric to frame, suturing and doubling fabric at the break as I had been taught. The result was a kintsugi-like effect, with the eye drawn immediately to the thread: in a crowd of rainy Paris or New York umbrellas, this might be the one you'd notice. But more importantly, it worked. (Apart from a slow drip under only the heaviest of rain conditions.) For the next several years this umbrella became my go-to, and I carried it through many storms and cities. I think about it and can picture it even now, while the countless other umbrellas of my lifetime have faded from memory. For a time, I was something other than a user to it, and it was something other than an object to me: something else or other had been sewn into the fabric with the thread. The umbrella fixer had a *working relation* to umbrellas – and for a while, before sadly leaving it in the overhead bin of a transcontinental flight, so did I.

*Story number two*: We live on a small farm in upstate New York. Bud, our immediate predecessor on the land, was a mechanic at our local university's particle accelerator. Beyond the house itself, full of strange and jerry-rigged fixes – mysteriously repurposed school bells, unmarked drawer knobs doubling as switches – the most distinctive feature of the place on our arrival was a John Deere 400-series tractor from the 1970s. A beast of a machine, it was the workhorse of the farm and survived many an adventure, going through the ice of the pond one winter, and by evidence on the casing, at least one internal engine fire (we added one more). In the summer it would regularly overheat and shut down. On cold fall days it smoked and generated enough heat to warm your face as you worked (pleasant but not recommended).

One spring it utterly failed to start and we called in a John Deere mechanic to diagnose. He was flummoxed: almost no parts were standard or original (it now included an engine from a different manufacturer altogether); components had been rebuilt and reordered in ways and for reasons hard to decipher; some parts, it appeared, had been remilled; and the basic functional logic of the machine had been utterly scrambled. And yet it worked – or had, until that spring. Now, without the human at the center of this unique and long-unfolding relation, the function of the machine was beyond saving, and we sadly sold it for parts. While he and it lived, our predecessor maintained a working relation with the tractor. We, and the John Deere mechanic, did not.

*Story number three (also from the farm)*: In the 1960s, Bud dug a pond. This served two purposes. The first was as source for the local volunteer fire department, who still come each year, practicing winter ice rescues and in spring drawing water into their trucks and shooting it back through the air to refill. The second was for hockey. When we moved onto the farm, we discovered and were mystified by a collection of more than thirty very beat up scraper shovels, the remnants (we later learned) of the unpaid child labor crew that would gather to clear the pond after a winter storm. Climatically, Bud's timing was excellent: the 1970s turned out to be one of the coldest periods on record in eastern North America, and the winter pond became a fixture of the neighborhood – we still encounter people decades later who know the house by the pond in its frozen and wintery form. Hockey pucks, with perhaps the longest half-life of modern sporting equipment, are still occasionally found in the woods.

When we took over the farm, we inherited as well, as I chose to interpret it, the moral responsibility of the pond. This involves an enormous and partly intentional amount of seasonal labor. Through this work – leaf removal, the watering of the orchard, the application of barley to aid in anaerobic decomposition, an extended experiment in hand dredging (!) to maintain depth – the pond has stayed clear, at a time when most other ponds around us, less artfully placed and balanced, have entered into a permanent eutrophic decline. The hardest and heaviest work however is reserved for winter, when the delicate dance becomes to remove the accumulating layer of snow before it builds to a weight that forces the ice down and the water up in ways that crystallize the accumulated snow, rendering it unusable for the winter. In some years the snow comes in before the ice freezes to a depth sufficient to bear human weight. In such years, or when temperatures fail to drop and stay sufficiently low, the pond is lost for the season and sits mournfully (as I regard it) through February drizzle waiting for spring. On glorious years where the cold comes early and stays, there remains the problem of surfacing, solved one year by a wildly inefficient system of micro flooding (if you drill enough small holes it turns out, the weight of the ice pushing down can cause the pond to *flood itself*); but more generally by a gas powered pump that on freezing winter nights can be used to draw water from the bottom of the pond to be sprayed via hose across the surface. In such years, friends, neighbors and kids from the preschool up the road will visit and skate, locating this relation in planetary time as well: I am conscious of the fact they may be the last generation to skate on a pond in upstate New York, an experience soon to fade from human and planetary memory. When a long-ago neighbor who had grown up skating on the pond returned recently to mourn and bury his mother, he skated on the pond each day. I recognized his and his adult children's right and importance to do so, and worked extra hard to keep the pond going amidst a season challenged by heavy snowfall and thaws. I hold these obligations – to Bud, to the kids, to the neighbor, to the pond itself – as deep and grounding commitments. In a way to be shortly explored, I hold them to be working relations.

My relation to the pond (and to the umbrella, and Bud's to the tractor) is a relation of use and function: I benefit from each, I use them purposively, and they help me to accomplish certain goals. The relation is also technical and epistemic, in the sense of involving detailed and specialized knowledge along with operations requiring some degree of skill, training or experience to perform. The relation is certainly one of labor, sometimes large amounts of it and inconveniently timed (the tractor breaks when it breaks; the winter storms and freezes come when they come). The relation also involves some element of puzzle and wonder: they remain partly mysterious to me and retain the capacity to surprise. And it can fairly be described as a relation of care, although I fear this partial and lossy description at once claims too much and too little. ('*Do you care for the pond?*' 'Yes: affectively and by action. If I were to lose it, I would mourn and my life would be diminished.' '*Do you care for the pond like you care for your kids?*' 'No not really. It is not that kind of thing.')

Function, labor, puzzle, care: all name elements, none are enough. We need new/old language, new/old words. Acts of maintenance and repair as explored in this volume live within and anchor different ways of being in and being with the worlds around us. They live within working relations.

**Seven Propositions**

First, a word on 'work' and its distinct usage and meaning in this piece. The notion of work deployed here does not follow and should not carry the connotation of payment, professionalism, or the bureaucratized or rationalized relations central to modern business, government and many forms of exchange. Nor does it put work on one side of a division between work and leisure, or related concepts like work/life balance; or accept the division of work into material vs. non-material vs. 'social' forms – for example, Hannah Arendt's (1958) three-way division of the human condition into labor, work and action. Rather, it is meant in the capacious sense of work coming out of the tradition of American pragmatist sociology: as for example when scholars like Susan Leigh Star would ask her students (would ask me) "*What work does this thing* [read: object, practice, infrastructure, belief] *do*?". Work then is about *the production of an effect,* whether a subtle force or change, or – at least as important and widely misunderstood – the quiet holding of things in place. When added to relation, this is meant to underscore two central points: first, that relations are not automatic, uncomplicated or for free; and second, that they are also never fully locked down, fixed or foreclosed, because the core of relations (at very least *some* relations) is work – and work is always ongoing. To put it glibly, work is always a work in progress. In this lies the hope and terror of the world.

What then are working relations?

Proposition one: *working relations are not flashy*. As anthropologist Kate Brown (2023) details in her account of the garden cooperatives of Tallinn under Soviet and post-Soviet rule, they involve a great deal of drudgery: the slow and heavy movement of earth, the tending and application of compost (outside the chemical fertilizer complex of the collective farms), the gathering and circulation of produce. Such efforts grew up as a kind of weed through the real and figurative cracks of the Soviet industrialisms of factory and farm. But such patterns hold elsewhere as well. Sometimes working relations make your feet hurt. Sometimes your eyes. Sometimes they involve the kind of work that nobody, including you, wants to do. As a mode of practice and a thing to study, as Leigh Star (1999) observed of infrastructure, working relations are often boring.

Proposition two: *working relations live at many scales*, from the relatively small and local to systems and infrastructures of larger reach and weight, including in their place-based and planetary entanglements. Working relations can, in principle, be artisanal or industrial in form: neither is automatically superior, and both forms matter. At the smallest level, as a philosophy of life and connection, working relations may be central to the kind of 'sympoesis' or 'becoming-with' described by Donna Haraway (2016) and exemplified by the relation of root to fungi which completes the circuit of tree to earth, passing elements – water, nitrogen, sugars – essential to each and which neither can independently source, produce or absorb. Importantly and as far as we can know, this remains a functional relation, organized around the mutual needs meeting of each side. At the same time, each actively cultivates the other, maintaining and repairing its counterpart in the service of an interrelated whole, or 'root mycorrhizal complex', that functions as the real and vital unit of analysis. At larger scales, working relations may best be understood not as properties of systems and infrastructures themselves, in any cybernetic or systems theory sense, but as a function of the worlds that grow at and around the cracks and intersection points of systems, rather like the matsutake worlds so beautifully and intricately described by Anna Tsing in *The Mushroom at the End of the World* (2015). And from Tsing, another observation: *working relations enmesh and sustain us in the collaborative work of survival*. As such, they are never pure. Working relations are contaminated. Working relations are contaminating.

Proposition three: as a philosophy and practice of objects, *working relations run against and beyond sleek and narrow ideals of 'usability'* so forcefully pushed and pursued by the tech industry and contemporary capitalist enterprise, without however giving up on the essential and unsentimental presence and role of function. To hold and maintain a working relation is to go beyond the aesthetics of consumption built around an abstracted concept or *idea* of the object – and an abstracted concept or idea of *ourselves* in relation to it – to hold oneself accountable to what for lack of a better term we might call, with some trepidation, the *thing itself*: not *a* tent (casual member of a general class), or *the* tent (representative of an ideal type) but *this* tent, here, right now, with its slight discolorations, the place the fabric has begun to sag, the seam imperfectly sealed that now admits water under certain conditions of rain and condensation. It may be the tent that has accompanied you through fondly remembered travels and holidays; or it may be the tent that keeps you alive, sheltered against the challenges, deprivations and hopes of an unhoused life.

Proposition four: As a philosophy of time, *working relations challenge the primacy of production and design, insisting instead on the long-run rhythms and unfoldings of things in the world*. This runs counter to the aestheticized snapshots of consumerist fantasy, addicted to demos, unveilings and unboxings. The time of working relations is sustained and rhythmic (Lefebvre 2004), embedded and committed through repetitions and refrains that lock human actors into other kinds of patterns and pacings – the temporalities of seasons, of growth and decay, of life and death. To abide by working relations is to accord oneself to the pace and rhythms of the world: its fastnesses and slownesses, its pauses and accelerations, its movements and repetitions at scales from the minute to the geological. Working relations are also *stretched* relations, relations that unfold and endure over time: as Bud's life unfolded into the tractor; as mine now unfolds into the pond. They take on different casts and characters, imprinting on worlds and bodies as they go. As the tractor aged, so did Bud, wrinkles deepening as the rust slowly built. Like many academics of my generation (and perhaps you too), I have an indent that has slowly grown behind the first knuckle of my middle finger (for me, this is on the right hand), the result of a decades long working relation with pen and paper. Future archaeologists may someday use this bone marker to identify me, correctly, as a member of the Late Dactylic.

Proposition five: *working relations require and build working knowledge*. As already indicated, the fixing and maintenance of umbrellas, tractors and ponds takes work, and that work frequently takes skill and expertise to execute well and thoughtfully (as well as a great deal of sometimes brute force labor). This working knowledge (Harper 1987; Crawford 2009) is both input and output of the activity: more accurately, it grows within and alongside the practices inscribed. This makes learning a – perhaps *the*? – central act of working relations, and the sharing of knowledge its most powerful mode of extension in the world. Conversely, efforts to limit or attenuate working knowledge – whether by design, strategy, or simple neglect and disregard – may undermine or destroy working relations, including those forms captured and expressed under the medium of ownership. Or as the intriguing claim from repair organization iFixit (n.d.) has it, "If you can't fix it, you don't own it." Relatedly…

Proposition six: *working relations may scramble and reconstruct our basic categories of property*. At their best, they make ownership reciprocal: under prevailing property relations, I own the pond, but in some real and effective sense, the pond owns me. It can place claims and demands; it can control my time and choices; it can impose hard obligations; it can make me do things I'd sometimes rather not do. This reciprocal character runs against the absolutist myth of ownership invented and imposed by John Locke (1690), itself the necessary precursor to enclosure, empire and colonial dispossession. Working relations remind us that against and beneath this world-breaking / world-making fantasy stands another possibility: that to hold a relation is to be held *by* and *in* relation, whether embrace or constraint, care or confinement (and quite possibly all these things at once).

Proposition seven: *working relations may include elements of reverence and wonder* that have been all too frequently erased from contemporary human relations with things (or at very least, our self-accounts of same). As Max Weber (2002 [1905]) famously observed in his account of the iron cage of modernity, the narrowing of relations to function and the rise of calculative rationality has fostered a progressive disenchantment with the world of things, elevating relations of convenience, thin use, and disregard over richer, older and other alternatives. These latter included at the margin spirit, and the kinds of wonder and mystery, the sense of a world-beyond-us-and-yet-part-of-us, associated with the divine. But like a cartoon character running through a wall, this has left a god-shaped hole in our thinking, about objects as about much else, that has been all too often and easily claimed by a certain kind of outgrowth or shell of conservative thought, now hollowed out and reclaimed by a moment and movement become radical and nihilistic.

One description of modernity – though we should never take such self-descriptions *too* seriously: there are many modernities and may be more still (Therborne 1995), we have never been as modern as we might suppose (Latour 1993), and who is the we in this sentence anyway? – is that it proceeds through the rupture and attenuation of working relations. The great transformation (Polanyi 1944) is a great disembedding, wherein relations are lifted up, emptied, and returned to earth as system lacking specificity, weight, and depth. The utter impossibility of this operation (per Latour's astute observations) does not prevent its power, and we occupy its aftermath. Through these processes people learn to engage with figures of things, rather than things themselves, to live in places without *living in places*, to be surrounded by things, and yet find themselves utterly and disorientingly alone. This can be and has been accomplished in many ways: the engineered emptiness of commodification and the commodity form (Marx 1967; Polanyi 1944); the kinds of alienation produced by the collapse of public and private life and the ascendance of the self as last solid ground of experience (Arendt 1958); perhaps even, per conservative critiques of liberalism (Deneen 2018; Kingsnorth 2025), the impossible loneliness of the rights bearing individual themselves.

But even Nietzsche carried an umbrella.[ii]

## Blueberries in Chernobyl

A final figure, one that has haunted and challenged me throughout the writing of this paper. At a recent talk, Kate Brown[iii] shared a picture and description of a girl picking blueberries in Chernobyl, a glancing and enigmatic image from fieldwork. As I recall it, the image, taken at some distance, shows the girl, perhaps 10, glancing (or staring?) back at the camera, face expressionless (or defiant?). In her hand is a plastic pail like the one you might see in a child's hand at a beach, its top almost overflowing with blueberries. "Almost", because the girl has also clearly been eating: her fingers but also lips are stained blue. Per Brown's account, this is the extent of the interaction: this is a chance encounter, not planned fieldwork, in the forest just beyond the Chernobyl exclusion zone (but well within the shadow of radioactive effect) which has now become a vibrant refuge for wildlife, biodiversity, and it seems, blueberries. The image and story feels mundane and fraught, particular and enigmatic but also planetary in scope and scale.

Strangely, the scene draws out a personal/planetary memory of my own. Approximately 1.8 billion years ago (scientists believe) the third largest known asteroid strike in Earth's history took place in what is now called, by some,[iv] the Sudbury basin in northern Ontario (but was then part of the as-yet unseparated supercontinent of Nuna). One-point-eight billion minus 90 years later, my grandfather, having lost his Manitoba farm in the North American Great Depression and despairing of the violence and transience of life as a horse logger in the lumber camps of northern Ontario, arrived as a nickel miner at the International Nickel Company (INCO), then the largest and most important site of nickel production in the world.

As a visitor to my grandparents' house in the 1970s and early 1980s, I spent long hours playing on the blackened rocks and hills of Sudbury. The vegetation, always thin, had by this time been almost entirely killed off by pollution coming from the slag heaps and the INCO stacks.[v] The result was a strange moonscape: empty, desolate, and utterly fillable by imagination. I spent hours exploring this world both strange and wondrous: I can still feel the hot touch of black summer rock on skin. In the winter, we would skate for miles on the too-clear (acidified) lake at the bottom of the street. Periodically we would encounter other kids doing the same and pause to discuss important news: for example, the story-shared-as-truth that Sudbury (and our hometown of Sault Ste. Marie) were the third-ranked targets on the Soviet bomber list, and thus that we were almost certain to be obliterated first – a ranking discussed solemnly and with some pride. (The question of how children in isolated northern Ontario towns might have access to secret Soviet bombing plans was never broached.)

I want to set these strange and eerie landscapes, Sudbury and the blueberry picker, against other ways of picturing and relating to landscapes, including ones indebted to a logic of purity and separation. I also mean to set them as limit cases and starting points for the kinds of working relations described here. These are powerful relations, redolent with meaning and peril. The Sudbury rocks meant (and mean) something to me, as I presume the blueberries do to the girl. These are not light or glancing, thin encounters in a casual world: they are not 'objects' in any narrow or simple sense. The blueberries that sustain the girl seem likely to someday kill her, as the nickel may someday kill me (and perhaps did kill my father, who died of a rare cancer around the age I am now as I write). At the same time there is a depth of relation that cannot be captured in the accounts of loss, degradation, and damage (Tuck 2009) that provide all too frequently our main or only line of approach to these scenes. These relations speak to the depth and capaciousness of humans, and to the depth and capaciousness of worlds. Readers of this piece may have Chernobyls or Sudburys of their own: places that on some level and for many reasons they shouldn't love, *but do*. Against romantic, narrow, and separated-out theories of Nature, or beliefs that long for purity as a refuge or starting point for change, the lesson is simple: *One can love a blasted landscape. One can love a broken world.*

But can we build *working relations* with such scenes and entities? Can we find new/old ways of being with such things that engage them in their plain and ordinary guises, their multiple rhythms and unfoldings, their multiple and nested scales? Can we do this in ways that move them (and us) beyond thin and depleted categories of property? Can we do all this with knowledge, with wonder, and without making knowledge the enemy of wonder? I do not know. But any practice of working relations must reckon with this complexity, this joy, and this violence.

**Conclusion**

It is difficult, perhaps impossible to think all the way to the bottom of our ideas. There is always a beyond we can just barely perceive, like a pearl glimpsed on the ocean floor before oxygen fails us and we must return, gasping and straining, to the surface.

Working relations is one poor effort at getting at the beyond of repair: to explain what is at stake (or *could* be) in reparative relationships with things and worlds; why fixing in its simple and mundane form might be linked to wider projects of endurance, reimagination and hope, even and especially amidst dark times; and why I loved my umbrella, Bud loved his tractor, and why I labor, endlessly and irrationally, over the pond.

Working relations are a plea for the present-at-hand over the powers and dangers of transparency;[vi] plain language over cipher; the virtues of rhythm and timefulness over timelessness (and its strange cousin speed). Working relations acknowledge the potentialities of care over more instrumental relations, but they do not reduce or circumscribe one to the other, or place affect and function in essential conflict (though they *can* be in conflict in any given situation): if a theory of care, it is a thingy and capacious one (Denis and Pontille 2025). Viewed from the standpoint of human actors, working relations may be described alternatively as a kind of thick and layered functionalism or a stuff-y psychology, a way of explaining anchorings and attachments that tie minds to things, and bodies to place and earth, and mind and body *to each other* and to the core, beyond and before the Cartesian fork that we have spent so much time suffering or repairing. Working relations may also provide a hedge against tendencies towards romanticized or aestheticized relations between people, places and things which under the right (wrong) conditions can produce powerful tropes and figures – like 'wilderness' or the imagination of a people-less 'Nature' (Cronon 1996) – that can have certain and terrible land-emptying (by people-erasing) effects. Finally, working relations are often (always?) compromised, messy and imperfect, offering guidance but little of the adjudicatory clarity that can seem so essential to the strident normativities of our day.

If I have given examples that are personal and near-to-hand, I believe this way of thinking may also be used to pose questions to technologies and infrastructures of other kinds and scales. Could working relations help us to rethink our connections to the industrial infrastructures that remain very much with us, in but also well beyond places like Sudbury and Chernobyl? Could they help us to rethink our relations to computing in ways that go beyond sad and obfuscating metaphors like 'the cloud'? Could we conceive (and build and maintain) a working relation to artificial intelligence ('AI') tools and systems, leaving material and cultural worlds richer rather than poorer in their wake – a relation which surely must look very different from the one we have arrived at to date?

The varied papers in this issue each offer instances and varieties of these propositions, played out across a range of practices, places and worlds. They call attention to the depth and complexities of working relations, and their varied and essential groundings in function, labor, puzzle and care: sometimes in harmony, sometimes in tension, never fully reducible one to the other. That the practices and objects are very often mundane and ordinary is their strength, not their weakness: we need more and better accounts of ordinary practices with ordinary objects by ordinary people, rendered in their depth and specificity.

When I think of working relations, I think of the radical specificity of place, how different people and things are in every place we encounter them, and how different but also partly the same it must have been to grow up in Estonia versus northern Ontario in the 1980s. Where did the children of Estonia rank on Western bombing lists, and were they proud? I think of the long cold war between people and things, and how our imaginations of both and the lines between them became so narrow, so brittle and so fixed.

I think too of time, and of working relations as an always-already relation. We grow out of them, we grow into them. "Tell me about the first time you saw the northern lights" is a question for southerners, outsiders and tourists (Arluk and Eira 2025). At their richest, they are relations individuals and communities hold with worlds that they recognize to precede them and outlive them, emerging from and disappearing into wonder and mystery. This is the opposite of a blank-slate empty-map relation, which sees human and non-human inheritances as a flat surface to be scraped, buried and replaced: often in the interests of power; always in the interests of forgetting.

I think finally of the umbrella, the winter pond and Bud's tractor. I think of his end, and I imagine mine. We come into the world fixing, and we leave when we put the tools down.

---

[i] Sustainability in Practice: DIY Repair, Reuse and Innovation, 30 October–2 November 2024, Tartu, Estonia.
[ii] The most famous discussion of Nietzsche's cryptic umbrella note comes in Jacques Derrida's *Spurs: Nietzsche's Styles* (1981 [1976]), which initiates a wide-ranging excursion through sexuality, memory, interpretation, philosophical style, and the limits of knowability. My suggestion here is simple and direct: practically and philosophically, Nietzsche has forgotten his umbrella.
[iii] Kate Brown, "Tiny Gardens Everywhere: The Past and Present of Urban Self-Provisioning", public lecture at Cornell University, October 24, 2025.
[iv] In Ojibwe, the language of the local Atikameksheng Anishnawbek and Wahnapitae First Nations, the land is known as N'Swakamok, which translates to 'where the three roads meet' reflecting the area's longstanding role as a trade route, including a trade in copper going back several thousand years.
[v] Years later this situation would improve with the construction of the INCO superstack, designed to blast pollutants higher into the atmosphere to settle over a wider catchment area, ultimately precipitating the acid rain crisis (and US–Canada acid rain treaty) of the 1980s.
[vi] Recalled now in its original and essential meaning of invisibility. I can think of few metaphors that have been more systematically abused and misapplied; glass, when doing its job, is precisely what you *can't* see.